\documentclass[english,twoside,a4paper,10pt]{article}

\usepackage[latin1]{inputenc}
\usepackage[T1]{fontenc}
\usepackage[english]{babel}
\usepackage{amsmath}
\usepackage{amsfonts}
\usepackage{graphicx}
\usepackage{subfigure}
\usepackage{a4wide}
\usepackage{amssymb}
\usepackage{fancyhdr}
\usepackage{mathrsfs}
\usepackage{color}
\usepackage[toc,page]{appendix}

\def\ba{\begin{eqnarray}}
\def\ea{\end{eqnarray}}

\def\be{\begin{equation}}
\def\ee{\end{equation}}

\begin{document}

\title{{\bf A note on non  singular Einstein-Aether cosmologies }} 

\author{ 
Alessandro Casalino$^{1, 2}$\footnote{E-mail address: alessandro.casalino@unitn.it},\,\,\,
Lorenzo Sebastiani$^3$\footnote{E-mail address: lorenzo.sebastiani@pi.infn.it},\,\,\,
  Sergio Zerbini$^1$\footnote{E-mail address: sergio.zerbini@unitn.it}\\
\\
\begin{small}
$^1$Dipartimento di Fisica, Universit\`a di Trento, Via Sommarive 14, 38123 Povo (TN), Italy
\end{small}\\
\begin{small}
$^2$TIFPA - INFN,  Via Sommarive 14, 38123 Povo (TN), Italy
\end{small}\\
\begin{small}
$^3$ Istituto Nazionale di Fisica Nucleare, Sezione di Pisa, Italy
\end{small}\\
\begin{small}
Dipartimento di Fisica, Universit\'a di Pisa, Largo B. Pontecorvo 3, 56127 Pisa, Italy.
\end{small}
}

\date{}

\maketitle
\abstract{An effective Lagrangian approach, based on an extended Einstein-Aether (EA) model, is proposed. This model is presented as an alternative to GR and mimetic gravity models in order  to deal with the Big-Bang  cosmological singularity issue. In particular, working on non flat  Friedmann--Lemaitre--Robertson--Walker (FLRW) space-time,  a generalized Friedmann equation is derived, and it is shown that, for a suitable choice of the action within the extended EA theory,  a regular bounce solution is present, generalizing the bounce solution  obtained in Quantum Loop Cosmology (QLC) in the flat case. Furthermore,  perturbation theory of the extended model is investigated, and the main perturbation quantities are evaluated and the conditions to avoid physical (superluminarity) and mathematical instabilities are discussed. Finally,  the Static Spherically Symmetric (SSS) case is also investigated. It is found that, with the same action proposed for the FLRW space-time,  a Schwarschild like solution with a correction is obtained.
}

\section{Introduction}

The theory of General Relativity (GR) with the presence of a suitable cosmological constant and the addition of cold dark matter, the so called $\Lambda$CDM model, describes remarkably well a large part of the history of the Universe, including the acceleration (or dark energy dominated) era. In fact, the $\Lambda$CDM model has been recently tested with high accuracy \cite{Plank1,Plank2}.  

However, GR admits solutions corresponding to singular space-times, namely metrics whose scalar curvature invariants have singularities, or, equivalently, geodesic incomplete metrics exist. For instance, in the cosmological context, the so-called Big Bang singularity occurs. But, with unconventional equation of state for the matter content, it is not difficult to propose models with bounce solutions at $t=0$. In fact, consider the (spatially) curved FLRW
\be
ds^2=-N^2(t)dt^2+a(t)^2\left(\frac{dr^2}{1-kr^2}+r^2 d\Omega_2^2  \right)\,,
\label{metricn}
\ee
where $a(t)$ is the scale factor, $N(t)$ is the lapse function, $k$ is the spatial curvature ($k=0$ is the flat case) and $d\Omega_2$ is the two dimensional sphere metric. If we reduce to the flat case, where $k =0$, the usual first Friedman equation is
\be
3H^2=\rho\,, \quad H(t)=\frac{\dot{a}}{a}\,,
\ee
and the matter conservation equation
\be
\dot{\rho}=-3(\rho+p)H\,,
\ee
where we consider the units $8\pi G_N=1$, the dot denotes derivatives with respect to the time $t$, $H(t)=\dot{a}/a$ is the Hubble factor, $\rho$ and $P$ are respectively the energy density and pressure of the matter content fluid. From the two equations above we can obtain the second Friedmann equation
\be
\dot{H}=-\frac{1}{2}(\rho+p)\,.
\ee
As a result, assuming an equation of state
\be
p=\omega \rho\,,
\label{es}
\ee
and using the Friedmann equations, we find
\be
H(t)=\frac{2}{3}\frac{1}{\int (1+\omega) dt}\,.
\ee
In the $\Lambda$CDM model, the equation of state parameter $\omega$ is constant and we obtain the GR singular Big Bang solution. However, assuming a time dependent $\omega$ we can obtain regular solutions. For example, considering (see for example \cite{Qui} and references therein),
\be
1+\omega(t)=c_1-\frac{c_2}{t^\alpha}\,,
\ee
we find
\be
H(t)=\frac{2}{3}\frac{(\alpha-1)t^{\alpha-1}}{c_2+(\alpha-1)c_1t^\alpha}\,,
\ee
where $c_1\,,c_2$ and $\alpha$ are fixed parameters. Consequently, a bounce solution exists when $c_1,c_2>0$ and $\alpha>1$, such that $H(0)=0$ and $\dot H(0)>0$. However, one should note that the singularity here is present at $t=0$ in the pressure $p$. More realistic cosmological bounces models have been studied, see for example \cite{staro15,odin20}. 

Another approach to solve the singularity issue is based on quantum corrections to GR. In fact, in the context of cosmology, an effective modified Friedman equation has been obtained in the so called Quantum Loop Cosmology theory, whose solution in a flat Friedmann--Lema\^{i}tre--Robertson--Walker  space-times admits a \textit{bounce}, a solution without the GR Big Bang singularity \cite{Bo,Bo2,Bo3,da19}.

In this paper we present a theory which avoids the drawbacks of the first approach, related to the singularity of matter observables, and provides the framework to extend QLC results to the curved case. We consider a specific class of modified Lorentz-violating gravitational models called Einstein-Aether models \cite{Ted,Gaspe}, whose implementation, similarly to mimetic gravity, can be performed with the addition of a Lagrange multiplier to the action. The original AE model contains four parameters which describe deviation from GR via the Aether vector field coupling with the metric. These parameters can be constrained using several experimental results, see for example references quoted in \cite{Ted1}. In the original model, the problem of initial singularity has not been fully solved (see Ref. \cite{Campista,Chan}). The EA models have also been studied in other works, see for instance Refs. \cite{ae1,ae2,ae3}.

In particular, we propose an extension of AE theory which can extend QLC bounce solutions for curved FLRW space-times, providing the QLC Friedmann equations in the flat case, and that can be easily reduced to the original AE model. Extended EA models have also been studied in several works, see for example Refs. \cite{TedDon, Zlosnik, Barrow12}.

The content of this paper is the following. Firstly, in Sec. \ref{sec:ext_mim} we review the standard mimetic gravity theories in order to show similarities and differences with respect to the EA theory. In Sec.\ref{sec:AE} we present a specific extended EA model, and show that its associated generalized Friedmann equations admit non-singular solutions. We also provide the first order perturbation equations and show the condition to avoid superluminarity and gradient instabilities. Finally, we study the model on a Static Spherical Symmetric (SSS) space-time, and provide the solutions for some choices of the extended AE action. In Sec. \ref{sec:conc} we draw the conclusions.

Note that both the original EA model and the mimetic gravity model, and their respective extensions, contain vector and scalar fields with fixed four norm. As already mentioned, in our approach this constraint is implemented using a Lagrangian multiplier approach. 

If not otherwise stated, in this paper we fix the convention $c=1$ and $8\pi G_N=1$.

\section{Extended mimetic gravitational model}
\label{sec:ext_mim}

In this section we provide a brief review of extended mimetic gravity, in order to show the similarities between these models and our proposed extended AE models. We follow Refs. \cite{Casa19,Seba}. The relevance of mimetic models consists in the fact is one of the theories in four dimensions which provides second order differential equations on FLRW space-time \cite{muk,muk1,vik,Derue,barvi,mata,golo,Lim,Capo,Odi1,Odi2,Odi3,myrza1,myrza2,myrza3,Rabo,mim_sv1,mim_sv2}, among with the Horndeski model \cite{Horn}, DOHST models \cite{Deffa,DeFe}, or Non Polynomial Gravity models \cite{aimeric,aimeric2}. In this review, by means of the Lagrangian mini-superspace approach within the non flat FLRW space-times in Eq. (\ref{metricn}), we study an extended mimetic model introduced in Ref. \cite{Muk3}.

The action reads
\be
I =  \int_{\mathcal M} d^4 x \sqrt{-g} \left\lbrace\frac{R}{2} + \lambda\left(X-\frac{1}{2}\right)+f[\chi(\phi)] \right\rbrace+I_m\,,
\label{d_m}
\ee
where $g$ is the determinant of the metric $g_{\mu \nu}$, $X \equiv -\frac{1}{2}g^{\mu \nu}\partial_\mu \phi \partial_\nu \phi$,  $\lambda$ is a Lagrange multiplier field, $\phi$ is the mimetic scalar field, and $I_m$ is the usual matter-radiation action. The higher order differential term in $\phi$, depends on $\chi(\phi)=-\nabla^\mu \nabla_\nu \phi \, /3$. With the metric \eqref{metricn}, the action is a functional of $a(t)$, $N(t)$ and $\lambda(t)$. We assume $\phi = \phi(t)$, i.e. an homogeneous field which depends only on $t$. We obtain the Lagrangian
\begin{equation}
L=-6\frac{\dot{a}^2 a}{N}+6kaN+Na^3f[\chi(\phi)]+N a^3 \lambda \left(\frac{\dot\phi^2}{2N^2}-\frac{1}{2}\right)+\frac{a^3(\rho+p)}{N}-Na^3(\rho-p)\,,
\end{equation}
where $\rho$ and $p$ are the energy density and pressure of matter and the dot denotes the derivative with respect to $t$. Furthermore we have 
\begin{equation}
\chi(\phi) = -\frac13 \nabla^\mu \nabla_\mu \phi  = \frac{H-\dot N/(3N)}{N^2}\dot\phi\,.\label{eq:chi}
\end{equation}

The variation with respect to $\lambda$ gives the so-called mimetic constraint
$X=1/2$, and therefore $\phi= t$. The variation with respect to the lapse function $N$, where we replace $N=1$ and $\dot\phi= 1$ after the computation, gives the generalized first Friedmann equation
\be
 6\left(H^2+\frac{k}{a^2}\right)+f(H)-H\frac{d f(H)}{dH}  =2\rho +2\lambda-\frac{1}{3}\frac{d}{d t}\frac{d f(H)}{d H} \,,
\label{z20}
\ee 
where $f$ is now a function of $H$, as we can see from Eq. \eqref{eq:chi} evaluated with $N=1$. The variation with respect to the field $\phi$ leads, after integration, to \cite{Muk3}, 
\begin{equation}
    \lambda=\frac{C}{a^3}+\frac{1}{6}\frac{d}{d t}\frac{d f(H)}{d H}\,,
\end{equation}
where $C$ is an integration constant mimicking dark matter contribution. In the following we fix $C=0$. Thus the first Friedmann equation \eqref{z20} becomes
\be
 6\left(H^2+\frac{k}{a^2}\right)+f(H)-H\frac{d f(H)}{dH}  =2\rho\,.
\label{z2}
\ee
Finally, the variation with respect to the $a$ gives the generalized second Friedmann equation,
\be
 3H^2+2\dot{H}+\frac{f(H)}{2}-\frac{H}{2}\frac{d f(H)}{dH}-\frac16 \frac{d}{dt}\frac{d f(H)}{dH}  =-p  \,.
\label{z1}
\ee

Deriving the Friedmann equation and making use of the above results, we get the matter conservation equation
\be
\dot{\rho}=-3H(\rho+p)\,.\label{c}
\ee
As a result, when $p=\omega \rho$ with $\omega$ constant, we obtain the well known solution  
\be
\rho(t)=\rho_0 a(t)^{-3(1+\omega)}\,.
\label{c1}
\ee

Consider the following choice for the arbitrary function $f$ \cite{Muk3,Lang,Haro},
\begin{equation}
 f(H)=6H^2+\frac{12}{\alpha^2}\left[
 1-\sqrt{1-\alpha^2 H^2}-\alpha H\arcsin\left(\alpha H\right)
 \right]\,,\label{eq:emp}
\end{equation}
where $\alpha$ is a dimensional positive parameter. Since $f(H)$ goes to zero when $\alpha\rightarrow 0$, in this limit we recover GR. Thus $f(H)$ may represent a ``correction'' to Einstein gravity.
The first Friedmann equation (\ref{z2}) with this choice of the function becomes
\begin{equation}
\frac{6}{\alpha^2}\left[
1-\sqrt{1-H^2\alpha^2}\right]=\rho -\frac{3 k}{a^2}\,,\label{EOM1bis}
\end{equation}
which is equivalent to 
\begin{equation}
3H^2=\left(\rho-\frac{3k}{a^2}\right)\left[1-\frac{(\rho-\frac{3k}{a^2})}{\rho_c}\right]\,,\qquad \text{where}\qquad  \rho_c=\frac{12}{ \alpha^2}\,.
\label{zf}
\end{equation}
An alternative Lagrangian derivation within a mimetic approach, with $k \neq 0$, is available in Ref. \cite{Lang} and references therein.

In the flat case, we obtain the QLC Friemann equation
\begin{equation}
3H^2=\rho\left(1-\frac{\rho}{\rho_c}\right)\,.
\end{equation}
For an equation of state $p=\omega \rho$, it admits a bounce solution. Here the critical density is given by $\rho_c$. Furthermore, in the case $\omega=-1$, namely $\rho=\rho_0$, the above equation admits a flat $k=0$ de Sitter solution. For other cosmological bounce solutions see Ref. \cite{Biswas1,Biswas_2} and references therein.

However, although mimetic models admit non-singular bounce solutions, they are plagued by gradient and/or ghost instabilities, see for example Refs. \cite{Ganz, Sunny}. This motivates an investigation of alternative models. In the following we consider an alternative with similar mimetic structure but different main tensor field rank. In particular, we replace the scalar field $\phi$ with a vector field. These models are the aforementioned extended AE models.

\section{The Einstein-Aether extended model}
\label{sec:AE}

In this section we propose an extended AE models defined on non flat FLRW. We denote by $u^\mu$ the time-like Aether 4-vector field. Analogously to the mimetic gravity case, where the evolution of the scalar field $\phi$ is fixed by the mimetic constraint via a Lagrange multiplier, this field norm will be constrained with a similar action addition.

The original and the extended model depend on the invariant
\be
\mathcal{K}=c_1 (\nabla_\mu u^\nu)^2+c_2(\nabla_\mu u^\mu)^2+c_3 \nabla_\mu u^\nu \nabla_\nu u^\mu+
c_4 u^\alpha u^\beta \nabla_\alpha u^\mu \nabla_\beta u_\mu \,. 
\label{k}
\ee
On a curved FLRW, the value of the invariant is $\mathcal{K}=3\beta H^2$ with $\beta \equiv c_1+3c_2+c_3$, where $c_{1,2,3,4}$ are adimensional quantities. Regarding the background field equations of motion, this justifies the replacement of $\mathcal{K}$ with another main variable defined as $\theta=-\nabla_\mu u^\mu=-3H$. In other words, $\mathcal{K}= \beta \theta^2/3$. In the following we consider a generic space-time with $\theta\equiv-\nabla_\mu u^\mu$, and then study specific space-times applications.

The action of the proposed AE extended model reads
\be
I =  \int_{\mathcal M} d^4 x \sqrt{-g} \left[ \frac{R}{2}+\lambda\left(u_\mu u^\mu+1 \right)-f(\theta) \right]+I_m\,,
\label{ae}
\ee
where $I_m$ is the matter action. With respect to the original work of Ref. {\cite{Ted}}, where the $f(\theta)=3\mathcal K=\beta\theta^2$ is fixed, in our extended approach the function can be generic. The equations of motions for the theory are
\begin{align}
    G_{\mu\nu} - \frac{1}{2} f(\theta) g_{\mu\nu} - \frac{1}{2} \lambda \left[2 u_\mu u_\nu + g_{\mu\nu} (1 + u_\rho u^\rho)\right]&\nonumber\\+\frac{1}{2}g_{\mu\nu} \left[\nabla_\rho u^\rho \frac{df(\theta)}{d\theta}+ u^\rho \nabla_\rho \nabla_\sigma u^\sigma \frac{d^2f(\theta)}{d\theta^2}\right] &= 0\,,\label{field}\\
    2 \lambda u_\mu - g_{\rho \sigma} \nabla_\mu \nabla^\sigma u^\rho \frac{d^2f(\theta)}{d\theta^2} &= 0\label{lambda}\,,\\
    1 + u_\mu u^\mu &=0\,,\label{u}
\end{align}
which was obtained varying the action respectively with respect to the metric $g^{\mu \nu}$, the vector field $u^\mu$ and the mimetic scalar field $\lambda$.

In the following sections we consider this theory on two different cases: the FLRW and static spherical symmetric space-times.

\subsection{Friedmann--Lema\^{i}tre--Robertson--Walker solutions}
\label{sec:AE_FLRW}

In this section we present the main results for a FLRW metric (\ref{metricn}). In order to find the equations of motion, we can proceed in two ways. One way is to evaluate the equations of motion \eqref{field}-\eqref{u}. We instead propose again the mini-superspace method which was applied to the extended mimetic model in Sec. \ref{sec:ext_mim}. 

We consider the time-like Aether 4-vector $u^\mu=(b,0,0,0)$, whose norm is given by $ u_\mu u^\mu=-N^2 b$. In the metric (\ref{metricn}) with $k \neq 0$, we have
\be
\theta=-\nabla_\mu u^\mu= -\frac{1}{\sqrt{-g}}\partial_{\mu}\left( \sqrt{-g}u^\mu \right)=
3H b+b\frac{\dot N}{N}+\frac{d}{d t}b
\,.
\ee
The related mini-superspace Lagrangian reads
\be
L=-6\frac{\dot{a}^2 a}{N}+6kaN-Na^3f(\theta)+N a^3 \lambda \left(1-N^2 b^2\right)+\frac{a^3(\rho+p)}{N}-Na^3(\rho-p)\,.
\ee
The variation with respect to $\lambda$ leads to $b=1/N$. Therefore the Aether field norm is $u_\mu u^\mu = -1$, is compatible with \eqref{u}. This confirms the goodness of our choice for the form of $u^\mu$.

Making the variation with respect to $N$, and then considering $N=1$ and $b=1$, we obtain the generalized Friedmann equation
\be
\frac{\theta^2}{3}+\frac{3k}{a^2}-\frac{f(\theta)}{2}+\frac{\theta}{2}\frac{d}{d \theta}
f(\theta)=\rho+\lambda-\frac{d}{2dt}\frac{d f(\theta)}{d \theta}\,, \quad \text{where } \theta=3\frac{\dot{a}}{a}=3H\,.
\label{F10}
\ee
The variation with respect to $b$ leads to,
\begin{equation}
    2\lambda=\frac{d}{dt}\frac{d f(\theta)}{d \theta}\,,\label{lambda_sub}
\end{equation}
such that equation (\ref{F10}) simply becomes
\be
\frac{\theta^2}{3}+\frac{3k}{a^2}-\frac{f(\theta)}{2}+\frac{\theta}{2}\frac{d}{d \theta}
f(\theta)=\rho\,, \quad \text{where } \theta=3\frac{\dot{a}}{a}=3H\,.
\label{F1}
\ee
Finally, the variation with respect to $a$ is the second Friedmann equation  
\be
\frac{2\dot{\theta}}{3}-\frac{2k}{a^2}=-(\rho+p)-\frac{1}{2} \frac{d}{dt} \frac{d}{d \theta}f(\theta)\,.
\label{F2}
\ee
From the equations above we can derive the matter conservation law
\be
\dot{\rho}=-3H(\rho+p)\,.
\label{cc}
\ee
In fact, this equation follows by deriving the first Friedmann equation \eqref{F10} with respect to the time $t$ and making use of the second Friedmann equation \eqref{F2}. We note that the matter conservation law is identical both in GR and in the mimetic extended model considered in Sec. \ref{sec:ext_mim}.

\subsubsection{Original model solution}

In this section we consider the choice 
\be
f(\theta)=\beta \theta^2\,
\ee
This is the same as considering the original model with $f(\theta)=3 \mathcal{K}$, as in Ref. {\cite{Ted}}. With this choice the first Friedmann equation reads
\be
3H^2\left(1+\frac{\beta}{2}\right)+3\frac{k}{a^2}=\rho\,.
\ee
We note that in order to obtain a positive matter energy density, we have to assume $\beta+2>0$. This is the Friedmann equation for the original AE model, whose solutions have been investigated in Refs. \cite{Campista,Chan}, where singular solutions have been found. In fact, as in GR, we can easily integrate the equation of motion when $k=0$. In this case, using the matter conservation law (\ref{cc}) with a barotropic matter fluid, i.e. a matter fluid with an equation of state $p=\omega\rho$ with $\omega$ constant, we obtain 
\be
\dot{\rho}=-\sqrt{\frac{6}{\beta+2}}(1+\omega)\rho^{\frac{3}{2}}\,.
\ee
The solution of this equation is
\be
\rho(t)=\frac{2(\beta+2)}{3(1+\omega)\,t^2}\,.
\ee
Therefore, similarly to GR, we obtain a Big Bang singularity at $t=0$. 

\subsubsection{A regular bounce solution}

In this section we present a regular bounce solution within the extended AE theory. The proposed function is similar to the one made in the extended gravitational mimetic model in Eq. \eqref{eq:emp}, namely
\be
f(\theta)=-\frac{2\theta^2}{3}-\frac{4}{3\alpha^2}\left[
 1-\sqrt{1-\alpha^2 \theta^2}-\alpha \theta\arcsin\left(\alpha \theta\right)
 \right]\,.
\label{z}
\ee
It should be noted that assuming the dimensional parameter $\alpha$ very small, one has
\be
f(\theta)=\frac{1}{18} \alpha^2 \theta^4+...\,.\label{expBModel}
\ee
Thus, with a small parameter $\alpha$, $f(\theta)$ represents a correction to GR which starts with a second order contribution in the invariant $\mathcal{K}^2$. 

Furthermore, with the $f$ function choice (\ref{z}), the first Friedmann equation is
\begin{equation}
3H^2=\left(\rho-\frac{3k}{a^2}\right) -\frac{3\alpha^2}{4}\left(\rho-\frac{3k}{a^2}\right)^2 \,,\label{aeb}
\end{equation}
where the critical density is defined as $\rho_c \equiv 4/3\alpha^2$. This is the same form of the modified Friedmann equation obtained in the extended mimetic gravity theory, Eq. \eqref{zf}. Again, for $k=0$ we obtain
\begin{equation}
3H^2=\rho\left(1-\frac{\rho}{\rho_c}\right)\,. \quad \label{aeb1}
\end{equation}
namely the QLC modified Friedmann equation in flat FLRW space-time. 

For $k=0$ is also possible to integrate the matter conservation law. In fact, writing $p=\omega \rho$, we obtain
\be
\dot{\rho}=-\sqrt{3}(1+\omega)\rho \sqrt{ \rho - \rho^2/\rho_c}  \,,
\ee
whose solution is      
\be
\rho(t)=\frac{\rho_c}{1+\frac{3\rho_c}{4}(1+\omega)^2 t^2}\,.
\label{pp}
\ee
We note that the GR limit is recovered when $\rho_c \rightarrow \infty$, or $\alpha \rightarrow 0$.

Using Eq. (\ref{pp}) and the matter conservation law, we find the known bounce solution for  $\omega\neq 1$,
\begin{equation}
a(t)=\left(\frac{\rho_0}{\rho_c}+\frac{3}{4}\rho_0(1+\omega)^2\,t^2\right)^{\frac{1}{3(1+\omega)}}\,,
\label{44}
\end{equation}
where $\rho_0$ is an integration constant.
Furthermore, in the case $\omega=-1$, namely $\rho=\rho_0$, the above equation (\ref{aeb1}) admits a $k=0$ dS solution.

On the other side, when $k \neq 0$, we should use Eq. (\ref{aeb}), which may be rewritten as
\begin{equation}
3 H^2=\rho-\frac{3k}{a^2}-\frac{\left(\rho-\frac{3k}{a^2}\right)^2}{\rho_c}  \,.
\end{equation}
In general we can show that the Big-Bang singularity at $t=0$ is absent. Firstly, we consider an example of exact solution. Consider the barotropic equation of state  $p=-\rho/3$, i.e. the equation of state parameter is $\omega= -1/3$, and therefore, $\rho(t)=\rho_0 a(t)^{-2}$. It is convenient to introduce the quantity $y(t)=a^2(t)$.
In this case, equation (\ref{aeb}) becomes
\begin{equation}
\frac{3}{4} \dot{y}^2=(\rho_0-3k)y -\frac{(\rho_0-3k)^2}{\rho_c} \,.
\label{zf1}
\end{equation}
The related solutions are
\be
y(t)\equiv a^2(t)=\frac{\rho_0-3k}{\rho_c}+\left(C \pm \sqrt{\frac{\rho_0-3k}{3}  }\,t   \right)^2\,,
\ee
where $C$ is an arbitrary dimensionless integration constant. Moreover we assume $\rho_0>3k$ in order to obtain a real solution. These are regular bounce solutions, with $a(0) \neq 0$. When $C=0$, the regular solutions become a unique symmetric bounce solution, namely
\be
a(t)^2= \frac{\rho_0-3k}{\rho_c}+\frac{\rho_0-3k}{3}  \,t^2  \,,
\ee
and the related density is also regular. When $\rho_c$ goes to infinity, we recover the GR solution, admitting the Big Bang singularity.

For a generic $\omega$, it is not easy to find an exact solution. Alternatively, we may start separating the variable in equation (\ref{aeb}) with $y=a^2$, namely
\be
\int \frac{dy}{\sqrt{Y(y)}}=t\,,
\ee
where we used the matter conservation law and defined
\be
Y(y)=\frac{4 \rho_0}{3}\left(y^{1/2-3\omega/2}-\frac{\rho_0}{\rho_c}\,y^{-(1+3\omega)}-\frac{3k\,y}{\rho_0} +
\frac{6k}{\rho_c}y^{-(1/2+3\omega/2)}-\frac{9k^2}{\rho_c\rho_0} \right)\,.
\label{zz}
\ee
If $k$ is not vanishing, the above integral can be solved analytically only for $\omega=-1/3$.

However, if we make an expansion around the critical point defined by  $ Y(y_*)=0$, namely
\be
y^{1/2-3\omega/2}_*-\frac{\rho_0}{\rho_c}\,y^{-(1+3\omega)}_*-\frac{3k\,y_*}{\rho_0} +
\frac{6k}{\rho_c}y^{-(1/2+3/2\omega)}_*-\frac{9k^2}{\rho_c\rho_0} =0\,,
\label{z3}
\ee
we can look for an approximate solution, valid for small $t$, which is given by
\be
y(t)\simeq y_*+\frac{Y'_*}{4}\,t^2\,,
\label{zz4}
\ee
where
\be
Y'_*=\frac{4 \rho_0}{3}\left(  \frac{1-3\omega}{2}y_*^{-1/2-3\omega/2}+\frac{\rho_0(1+3\omega)}{\rho_c}\,y^{-(2+3\omega)}_*-\frac{3k}{\rho_0} -
\frac{3k(\frac{1+3\omega}{2} )}{\rho_c}y^{-(3/2+3\omega/2)}_* \right)\,.
\label{zzz}
\ee
In the above equation, $y_*$ is the solution of the transcendental equation (\ref{z3}).  It is easy to show that for $\omega= -1/3$, we get the approximate solution related to the exact solution found before. Moreover, if $k=0$ we recover $y_*=(\rho_0/\rho_c)^{2/(3(1+\omega))}$, such that
\begin{equation}
y(t)\simeq \left(\frac{\rho_0}{\rho_c}\right)^{\frac{2}{3(1+\omega)}}+\frac{1}{2}\rho_0(1+\omega)\left(\frac{\rho_0}{\rho_c}\right)^{\frac{-1-3\omega}{3(1+\omega)}}t^2\,,    
\end{equation}
which is consistent with the result in Eq. (\ref{44}).

Thus, from Eq. (\ref{zz4}) we can conclude that we obtain a regular symmetric bounce when $y_*>0$ and  $Y'_*>0$. In the limit $\rho_c\rightarrow \infty$, with parameters $k=0\,$ and $\omega\neq -1$, we find $y_*=0$ and the Big-Bang singularity appears.

\subsubsection{Cosmological perturbation theory}

In this section we present the cosmological perturbation theory results in extended AE models within curved FLRW space-times, which has been also investigated in Refs. \cite{Battye,Baker}. We firstly study the scalar perturbations, and we refer to the perturbed FLRW metric in Newtonian gauge
\be
ds^2=-\left[1+2\Psi(t,\vec{r})\right]dt^2+a(t)^2\left[1+2\Phi(t,\vec{r})\right]\left(\frac{dr^2}{1-kr^2}+r^2 d\Omega_2^2  \right)\,,
\ee
where $\Psi$ and $\Phi$ are the non-homogeneous and non-isotropic Newtonian potentials. Moreover we choose to split the perturbation of the Aether 4-vector as
\begin{equation}
    \delta u^\mu = (\delta_u^0,\delta u^i).
\end{equation}

We notice that the perturbation of $\theta$ and $\mathcal{K}$ at first order are not the same. In fact the perturbation of $\theta^2$ at the first order is
\begin{equation}
    \delta \theta^2 = 18 H^2 \delta u^0 + 6 H \delta \dot{u}^0 + 6 H \dot{\Psi} + 18 H \dot{\Phi} \,,
\end{equation}
where $\delta u_0$ is the perturbation of the first component of the Aether 4-vector field. On the contrary, the first order perturbation of $3 \mathcal{K}$ is given by
\begin{equation}
    3 \delta \mathcal{K} = 18 \beta H^2 \delta u^0 + 18 c_2 H \delta \dot{u}^0 + 18 c_2 H \dot{\Psi} + 18 \beta H \dot{\Phi}\,.
\end{equation}
Since at the background level we identified $3 \mathcal{K} = \beta \theta^2 = (c_1 + 3c_2 + c_3) \theta^2$, we can conclude that the theory formulated in terms of $\mathcal{K}$ and $\theta$ are equivalent at the first order perturbation level only if we consider $c_1 + c_3 = 0$, i.e. $\beta = 3 c_2$.

We can now proceed to perturb the theory \eqref{ae}. At first order in the perturbations, the mimetic constraint \eqref{u} on the FLRW metric becomes the constraint
\begin{equation}
    \delta u^0 = - \Psi\,.\label{constraint}
\end{equation}
From the spatial part of the equations of motion \eqref{lambda} we find the relation
\begin{equation}
    \dot{H} \frac{d^2 f}{d\theta^2} \delta u_i = \left[\partial_i \dot{\Phi} - H \partial_i \delta u^0 \right] \frac{d^2 f}{d\theta^2}\,,
\end{equation}
where we substituted $\Psi$ and $\lambda$ using respectively the constraint \eqref{constraint} and the equation \eqref{lambda_sub}. This equation can be used to find $\delta u_i$ once we know $\delta u^0$ and $\Phi$.

If we consider no anisotropic stress (no perturbation of the $T_i^j$ components of the standard matter stress-energy tensor), we obtain an equation for the evolution of the $\delta u^0$ perturbation which is in the form
\begin{equation}
    \delta \ddot{u}^0 + \dots \delta \,\dot{u}^0 - \left(\frac{c_s^2}{a^2} \right) \nabla \delta u^0 + \dots \,\delta u^0 = f(\delta \rho, \delta P)\,,
\end{equation}
where the dots are different cofficients which depend of background quantities ($H$, $\rho$, ...), while $f$ is a function of the perturbations of the energy density $\delta \rho$ and pressure $\delta P$. The velocity $c_s^2$ is given by
\begin{equation}
    c_s^2 = \frac{1}{1-\frac{3}{4} \frac{d^2 f}{d\theta^2}}\,.
\end{equation}
In general, to avoid gradient instabilities, we require this quantity to be positive. Moreover, $c_s^2<1$ in order to avoid superluminarity. This means that $f(\theta)$ should satify
\begin{equation}
    \frac{d^2 f}{d\theta^2} < 0 \qquad \text{for any value of $\theta$}\,.
\end{equation}

We have also generalized the propagation speed of gravitational waves to non flat FLRW space-times. The result is
\be
c^2_T=\frac{1}{1 +(c_1+c_3)\frac{d f}{d \mathcal{K}}}\,,\label{eq:ct2}
\ee
which in our case, where $c_1 + c_3 = 0$, reduces to $c^2_T=1$. Therefore, every model which uses $\theta$ instead of $\mathcal{K}$ will always satisfy $c_T=1$ regardless of the choice of $f(\theta)$.

\subsection{Static Spherical Symmetric solutions}

In this section, we briefly investigate the existence of Static Spherical Symmetric  solutions in a specific extended AE models. We also provide the explicit solutions for particular choicea of the function $f$.

We consider the SSS space-time in the following form,
\begin{equation}
ds^2=-A(r)^2B(r)dt^2+\frac{dr^2}{B(r)}+r^2 d\Omega_2\,,
\end{equation}
where $d\Omega_2$ is the two dimensional sphere metric, and $A\equiv A(r)\,,B\equiv B(r)$ are functions of the radial coordinate $r$ only. By assuming $u^{\mu}\equiv u^{\mu}(r)$, the Lagrange multiplier constraint (\ref{lambda}), among with the field equations (\ref{field}), gives $u^{\mu}=(0, \sqrt{-B},0,0)$. Therefore, in general the vector field is imaginary on SSS space-times, similarly to the static case of mimetic gravity, see for example Ref. \cite{Sebastrenzo}. Thus we have
\begin{equation}
\theta^2=-\frac{1}{4r^2} \frac{(2rA'/A +2B+rB')^2}{B} < 0\,.    
\end{equation}

As mentioned before, for simplicity we restrict our AE static models to depend only on the scalar $\mathcal{K}=-\theta^2$, $\theta$ being an imaginary quantity. Furthermore, it is convenient to work with $\theta$ instead of $\mathcal{K}$, but with $f(\theta)$ a real quantity.
 
From \eqref{field}-\eqref{lambda} we obtain the equations of motion
 \begin{align}
    1 - B(r) - r \frac{d B}{d r} - r^2 f+ r \sqrt{- B(r)} \, \frac{d f}{d \theta} \left( 2 + \frac{r}{A} \frac{d A}{d r} + \frac{r}{2 B} \frac{d B}{d r}\right) + \nonumber\\
    + \frac{d^2 f}{d \theta^2} \left[2 B - r \,\frac{dB}{dr} + \frac{r^2 B}{A^2}\left(\frac{d A}{d r}\right)^2 - \frac{r^2}{2 A}\frac{d A}{d r}\frac{d B}{d r} + \frac{r^2}{4 B}\left(\frac{d B}{d r}\right)^2-\frac{r^2 B}{A}\frac{d^2 A}{d r^2}-\frac{1}{2} r^2 \frac{d^2 B}{d r^2}\right]=0\,,\label{SSSEOMs}\\
    1 - B(r) - 2 r \frac{B}{A}\frac{dA}{dr} - r \frac{dB}{dr}- r^2 f + r \sqrt{- B(r)} \, \frac{d f}{d \theta} \left(2 + \frac{r}{A}\frac{dA}{dr} + \frac{r}{2 B}\frac{dB}{dr}\right) = 0\,.
 \end{align}
In the following we present two examples with different $f(\theta)$ function choices.

\subsubsection{Linear case}

In this first example we consider the Schwarzschild gauge with $A(r)=1$, and $f$ linear in $\sqrt{\mathcal{K}}$, i.e. $f(\theta)=\gamma \theta$.
We investigate this model to prove the goodness our formalism. In fact, this model corresponds to the additive term  $\sqrt{\mathcal K}$, which by definition is a divergence of a 4-vector. Therefore we expect this choice to be a trivial correction to GR.

With this choice, the equations of motions reduce to the single equation
\be
 1 - B(r) - r \frac{d B}{d r} - r^2 f+ r \sqrt{- B(r)} \, \frac{d f}{d \theta} \left( 2  + \frac{r}{2 B} \frac{d B}{d r}\right)=0\,.
\ee
Since
\begin{equation}
\theta = - \frac{4 B + r \frac{dB}{dr}}{2 r \sqrt{- B}}\,,
\end{equation}
and $f(\theta)=\gamma \theta$, the last two terms cancel and one has
\be
 1 - B(r) - r \frac{d B}{d r}=0\,.
\ee
Therefore, the exact solution is given by the Schwarzschild solution
\begin{equation}
B(r)=\left(1-\frac{C}{r}\right)\,,    
\end{equation}
where $C$ is a mass term.

\subsubsection{Quartic case}

In this section we consider a quadratic model in $\mathcal{K}$. In particular, the function $f(\theta)$ is given by the first term of (\ref{expBModel})
\begin{equation}
    f(\theta)=\frac{1}{18} \alpha^2 \theta^4\,,
\end{equation}
which is the first term of the expansion of Eq. (\ref{z}), discussed in Section 2. In fact, at the zeroth order in $\alpha$, the model coincides with GR, thus  admitting the Schwarzschild solution. At the second order in $\alpha$ one may look for a solution in the form,
\begin{equation}
A^2=1+\alpha^2 \tilde A (r)\,,\quad B(r)=1-\frac{C}{r}+\alpha^2\tilde B(r)\,,
\end{equation}
where we consider $\tilde A$ and $\tilde B$ as perturbations of the main functions. Evaluating the first equation of (\ref{SSSEOMs}) for this choice, we find that the second order contribution in $\alpha$ of the equation does not depend on $\tilde A(r)$. This simplifies the equation and provides the solution for $\tilde B(r)$
\begin{equation}
\tilde B(r)=\frac{C_0}{r}+\frac{1}{96 r}\left[
\frac{81 C^2}{r^3}-\frac{261 C}{r^2}+\frac{249}{r}+\frac{3}{r-C}
+\frac{4}{C}\log\left(\frac{r}{r-C}\right)
\right]\,,
\end{equation}
where $C_0$ is a new integration constant.

Furthermore, using the second equation in (\ref{SSSEOMs}) at the second order, we obtain the expression for $\tilde A(r)$ through the solution for $\tilde A'(r)$,
\begin{align}
 \tilde A(r) = &\frac{1}{96 C}\left\lbrace\frac{3 C^2}{2 (r-C)^4}+\frac{3(3+C)}{(C-r)^2}+\frac{44-19 C}{C (C-r)}+\frac{108 C (5+C)}{r^3}\right.\nonumber\\
 &+\frac{54-72 C}{r^2}+\frac{90-111 C}{C r}+\frac{C (16+3C)}{3(r-C)^3}+\nonumber\\
 &-\frac{2}{C^2 (r-C)}\log \left(\frac{r}{r-C}\right)\left[ 23 r(2 C-1) + C (27-44 C) + 2 (C-r)\log \left(\frac{r}{r-C}\right)\right]\bigg\rbrace\,,
\end{align}
where we consider $C_0=0$. We note that these functions are badly divergent for $r \rightarrow C$, and their regime of validity is in the limit $\alpha^2 C\ll r$, for which we recover the corrections to GR. Thus, these results cannot be used to find corrections to the GR horizon. Moreover, we note that, for $r\rightarrow \infty$, we obtain $\tilde A\,,\tilde B\rightarrow 0$, i.e. the large $r$ asymptotically flat GR limit.

\section{Conclusions}\label{sec:conc}

In this paper we have proposed an effective Lagrangian approach based on an extended Einstein-Aether (AE) model in a generic spatially non flat FLRW space-time. After showing the similarities between the AE model and the mimetic gravity, which are respectively based on the presence of an additional vector and scalar, whose norms are fixed,  we have investigated an extended AE model, based on the presence of a generic function depending on the  invariant $\mathcal{K}$. 

Furthermore, making a suitable choice for the action, namely the Einstein-Hilbert one with the addition of non polynomial $f(\mathcal{K})$,  a generalized Friedmann equation has been obtained. The additional term may be interpreted as an effective quantum correction, inspired by Quantum Loop Cosmology and depending on an arbitrary parameter $\alpha$. 

This generalized Friedmann equation admits a non-singular bounce solution at $t=0$.  Furthermore, we have shown that this non-polynomial contribution, for small $\alpha$, leads to a correction to GR which is of order $\theta\sim (H)^4$ and becomes negligible at small curvature. 

We have also investigated the cosmological perturbations of the model, providing the necessary conditions to avoid superluminarity and gradient instabilities.

We have also studied the Static Spherically Symmetric solutions. We have found that the Schwarzschild solution can be recovered with a suitable choice of Lagrangian. 

Finally, the  correction to GR given by $f(\theta)\sim \alpha^2\theta^4$, namely the aforementioned small $\alpha$ limit used in the FLRW cosmological case, has been considered. As a result, small corrections to Schwarzschild solution, valid only for large $r$, have been presented.  

\subsection*{Acknowledgments}

This work has been partially performed with the Mathematica extensions \texttt{xAct} \cite{xAct} and \texttt{xPand} \cite{Pitrou:2013hga}. A. Casalino acknowledges the financial support of the Italian Ministry of Instruction, University and Research (MIUR) for his Doctoral studies.

\end{document}